\documentclass[floatfix,12pt]{iopart}
\usepackage{iopams}
\usepackage{graphicx}
\usepackage{url}
\usepackage{color}

\begin{document}

\title[Finite size corrections]{Finite size corrections to random boolean networks}

\author{Michele Leone$^1$, Andrea Pagnani$^1$, Giorgio Parisi$^2$,  and Osvaldo Zagordi$^3$}

\address{$^1$ ISI Foundation,Viale S. Severo 65, I-10133 Torino - Italy}
\address{$^2$ Dipartimento di Fisica, Universit\`a di Roma {\em La Sapienza}, P.le Aldo Moro 2, I-00185 Roma - Italy}
\address{$^3$ International School for Advanced Studies SISSA-ISAS, Via Beirut 2-4, I-34014 Trieste  - Italy}
\ead{zagordi@sissa.it}
\begin{abstract}
Since their introduction, Boolean networks have been traditionally
studied in view of their rich dynamical behavior under different
update protocols and for their qualitative analogy with cell
regulatory networks. More recently, tools borrowed from statistical
physics of disordered systems and from computer science have provided
a more complete characterization of their equilibrium behavior.
However, the largest part of the results have been obtained in the
thermodynamic limit, which is often far from being reached when
dealing with realistic instances of the problem. The numerical
analysis presented here aims at comparing - for a specific family of
models - the outcomes given by the heuristic belief propagation
algorithm with those given by exhaustive enumeration.  In the second
part of the paper some analytical considerations on the validity of
the annealed approximation are discussed.
\end{abstract}


\section{Introduction}

Boolean Networks (BNs) are dynamical models originally introduced by
S. Kauffman in the late 60s \cite{Kauf69}. Since Kauffman's seminal
work, they have been used as abstract modeling schemes in many
different fields, including cell differentiation, immune response,
evolution, and gene-regulatory networks (for an introductory review
see \cite{GCK} and references therein). In recent days, BNs have
received a renewed attention as a powerful scheme of data analysis and
modeling of high-throughput genomics and proteomics experiments
\cite{IlyaBook2005}.

The bottom-line of previous research has been the description and
classification of different attractor types present in BNs under
deterministic parallel update dynamics
\cite{Kauf69,Kauf1,samuelsson,BastollaParisi}. Special attention was dedicated to
so-called critical BNs \cite{Derrida} situated at the transition
between ordered and chaotic regimes. Recently, a new point of view has
been introduced that casts the original dynamical problem into a {\em
constraint satisfaction problem}, that can be studied with theoretical and
algorithmic tools inspired from the statistical mechanics of
disordered systems \cite{MPZ,Libro_Martin}.

Following this new approach presented in \cite{letter,jstat} it is
possible to study the organization of fixed points in the
thermodynamic limit in random Boolean networks. This leads to the
identification of the sudden emergence of a computational core, whose
existence is a necessary (but not sufficient) condition for a globally
complex phase where all fixed points are organized in an exponential
number of macroscopically separated clusters. This phenomenon is found
to be robust with respect to the choice of the Boolean functions, and
missing only in networks where all boolean functions are of {\sf AND}
or {\sf OR} type. In addition, the size of the complex regulatory
phase is found to grow with the number $K$ of inputs to the Boolean
functions.

The main motivation for this work is to check the robustness of the
above-reported results in networks of finite size, both from the
theoretical and algorithmic point of view. We aim at understanding how
reliable the predictions of message passing algorithms such as Belief
Propagation (BP) are on graphs of finite-size.

We performed an extensive numerical check on the
number of solutions in small samples, comparing BP predictions to
exact exhaustive enumeration algorithms results. Results of BP on
directed BNs are found to be in very good agreement with exact ones
when counting the expected number of fixed points. A worse
performance of BP is found in its prediction of local quantities such
as the single nodes marginal probabilities. Nevertheless, the
performance seems to increase with system sizes for already moderate
size values.

In \sref{MODEL_DEF} we define the model, in \sref{sec:meth} we
describe the Belief Propagation algorithm and the strategy we adopted
to obtain exhaustively all solutions of our BN's model. In
\sref{sec:num} we quantify numerically the agreement between BP and 
the exhaustive search results. Furthermore, a brief analysis on
the distribution of the overlap of the solutions is presented.
Finally in \sref{sec:theo} we present the analytical computation of
the first and second moment of the average number of solutions of our
model of BNs. Numerical solutions of the BP equations
strongly suggest that the
annealed calculation of the entropy (logarithm of the number of BN
fixed points) is exact in the large size random case. Fluctuations
over the annealed value are computed numerically and successfully
confronted with analytical estimates of the second moment of the
number of fixed points.  The large size limiting value of the
fluctuations around the annealed result is also computed, and some
considerations are discussed.

\section{Definition of the model}\label{MODEL_DEF}

\begin{figure}[htb]
\vspace{0.2cm}
\begin{center}
\includegraphics[width=0.75\columnwidth]{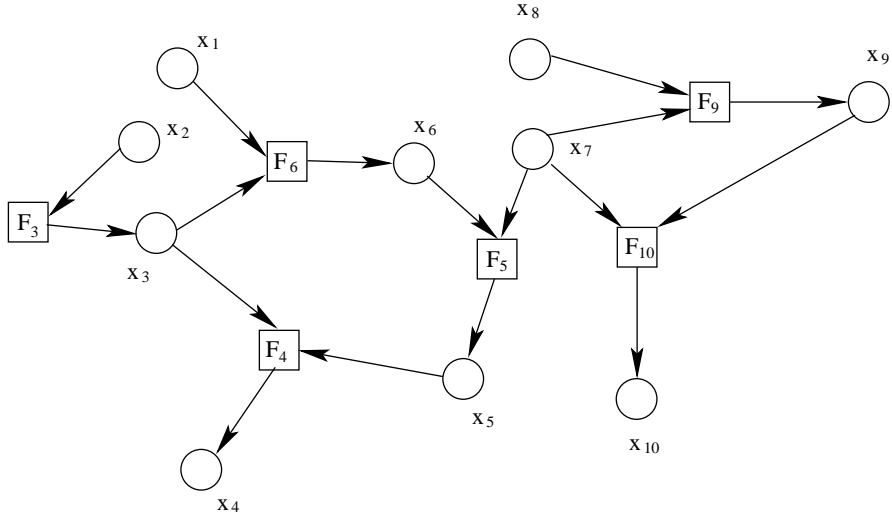}
\caption{Factor graph representation of a small Boolean network:
circles symbolize the variables, squares the Boolean functions. Arrows
stress the directed nature of the graph.  Variables $x_1,x_2,x_7,x_8$
are external inputs (non-regulated variables).}
\end{center}\label{fig:network}
\end{figure}

In the following we will consider a BN of $N$ interacting variables
$x_j\in\{0,1\}$ with $j\in\{1,\ldots,N\}$. In general, each variable
can be regulated by $K$ other {\em parent variables}, and can enter
in the regulation of an arbitrary number of {\em child variables}. We
then consider $M$ Boolean functions, represented by $F_a$ with
$a\in\{1,\ldots,M\}$, depending on $K=2$ inputs and having a single
output. The truth value of a given output variable $x_a$ is then fixed
by the truth values of the regulating variables $x_{a_1},x_{a_2}$ via
the relation:
\begin{equation}
\label{eq:function}
x_a = F_a ( x_{a_1},x_{a_2})
\end{equation}
with $a\in A\subset\{1,\ldots,N\}$ running over all regulated genes.
As shown in Fig.~\ref{fig:network}, not all variables need to be
controlled by a Boolean function, {\em i.e.}~in general we have
$|A|=M$ with $0\leq M\leq N$.
On the other hand, each regulated variable is the output of one and
only one function.  The whole set of $M=\alpha N$ Boolean constraints
completely specifies the network topology. We aim at computing
${\cal N}_\mathrm{sol}$, {\em i.e.} the number of stationary points of
the network. We introduce a
Hamiltonian, that it is equal to the number of unsatisfied Boolean constraints:
\begin{equation}
\label{eq:H}
{\cal H} = \sum_{a\in A} x_a \oplus F_a(x_{a_1},x_{a_2}) \equiv
\sum_{a\in A} E_a(x_a,x_{a_1},x_{a_2}) .
\end{equation}
where the symbol $\oplus$ stands for the logical operation {\sf XOR}.

We will consider {\em random factor graphs} characterized (among all possible random graphs) by: 
\begin{itemize}
\item[({\em a})] Function nodes $F_a$ have fixed in-degree
$K$ and out-degree one. 
\item[({\em b})] Variables $x_a$ have in-degree at most one. This
means that all regulating variables are collected in one single
constraint $F_a$ (see Eq.~(\ref{eq:function})).
\end{itemize}
Setting $\alpha := M/N$, the degree distribution of variable nodes
approaches asymptotically
\begin{eqnarray}
\rho^{\mathrm{out}}(d_{\mathrm{out}}) &=& e^{-2 \alpha }
\frac{(2 \alpha)^{d_{\mathrm{out}}}}{d_{\mathrm{out}}!} \nonumber \\
\rho^{\mathrm{in}}(d_{\mathrm{in}}) &=& \alpha \delta_{d_{\mathrm{in}},1}
+ (1-\alpha)\delta_{d_{\mathrm{in}},0}
\label{eq:degdistr}
\end{eqnarray}

We now have to specify the functions in the factor nodes present on the random factor graph 
defined so far, i.e we have to specify not only the topology of the graph, but also the content. There are $2^{2^2}=16$ possible Boolean functions with
2 inputs and 1 output. Following \cite{Kauf2}, we group them into 4
classes:
\begin{itemize}
\item The two constant functions equal to 0 and 1.
\item  Four functions depending only on one of the two
inputs, {\em i.e.} $x_1,\overline x_1, x_2,\overline x_2$. 
\item {\sf AND-OR} class: There are eight functions, which are given
by the logical {\sf AND} or {\sf OR} of the two input variables, or of
their negations. These functions are {\it canalizing}. If, e.g., in
the case $F(x_1,x_2)=x_1 \wedge x_2$ the value of $x_1$ is set to
zero, the output is fixed to zero independently of the value of $x_2$.
It is said that $x_1$ is a {\it canalizing variable} of $F$ with the
{\it canalizing value} zero.
\item {\sf XOR} class: The last two functions are the {\sf XOR} of the
two inputs, and its negation. These two functions are not canalizing,
whatever input is changed, the output changes too.
\end{itemize}

For the sake of clarity we concentrate only on classes depending
effectively on two inputs, {\em i.e.}~on those in the {\sf AND-OR}
class and the {\sf XOR} class. As we will see in the following
sections, the average number of  fixed points does not depend on the 
relative appearance of the functions within each class, but only on
the relative appearance of the classes. We therefore require $XM$
functions to be in the {\sf XOR} class, and the remaining $(1-X)M$
functions to be of the {\sf AND-OR} type, with $0\leq X \leq 1$ being
a free model parameter. In this simple case the network ensemble is
then completely defined by $\alpha$ and $X$.

\section{Methods: Belief-Propagation vs. Exact Enumeration}
\label{sec:meth}
Belief propagation is a marginalization algorithm used with success in
many different fields. It is exact on a tree, {\em i.e.} on a graph
with no loops, but it is also known to give good estimates of the
marginals in other cases.

In this section we will first introduce the BP algorithm, and then we
will explain the strategy adopted for the exhaustive search.

\subsection{The BP algorithm}
The BP algorithm \cite{Yedidia,SumProd,BMZ} is an iterative strategy
for computing the fraction of zero-energy configurations having, say,
$x_i=0$. More generally, BP calculates marginal probability
distributions of solutions of problems defined on factor graphs. A more detailed
introduction on the BP equations in the case of BNs can be found in
\cite{jstat}.

Let us define $a$ one of the clauses where $x_i$ appears. One can
introduce the following quantities:

\begin{itemize}
\item $\mu_{i\rightarrow a}(x_i)$: the probability that variable $i$
takes value $x_i$ in the absence of clause $a$.
\item $m_{a \rightarrow i}(x_i)$: are a function proportional to the probability that
clause $a$ is satisfied when variable $i$ takes value $x_i$.
\end{itemize}
The above-defined quantities satisfy the following set of equations:
\begin{eqnarray}
\label{eq:bp}
m_{a\rightarrow i}(x_i)  &=& \sum_{\{x_l\} \in a \setminus i} 
\left[1-  F_a (\{x_l\}_{l\in a})\right] 
\prod_{b \in a \setminus i} \mu_{b\rightarrow i} (x_i) 
\nonumber\\
\mu_{i\rightarrow a}(x_i) &=& C_{i\rightarrow a} \prod _{b\in i
  \setminus a } m_{b\rightarrow i}(x_i)\,\,\, ,
\end{eqnarray}
where $a \setminus i$ represents the set of all variable indexes
belonging to function $a$ {\em except} variable index $i$, $i
\setminus a$ the set of all function indexes to which variable $x_i$
belongs, {\em except} index $a$, and $C_{i\rightarrow a}$ is a
constant enforcing the normalization of the probability distribution
$\mu_{i\rightarrow a}$.  This set of equations can be numerically
iterated up to their fixed point, if present. It turns out \cite{jstat} that, in
the case of our model, the iterative procedure always converges.
Marginals can be computed in terms of the messages in
Eq.~(\ref{eq:bp}) via the following set of relations:
\begin{eqnarray}
\label{eq:marginals}
P_{a}(  \{x_i\}_{i\in a} )  &=& c_{a}\left[1- 
E_a(\{x_i\}_{i\in a})\right] \prod_{i \in a } 
\mu_{i\rightarrow a}(x_i)\nonumber \\
P_i ( x_i )  &=& c_i \prod_{a\in i}  m_{a\rightarrow i}\,\,\, ,
\end{eqnarray}
where $c_a$ and $c_i$ are again normalization constants.  From the
marginals we can eventually compute the entropy, {\em i.e.} the logarithm of 
the number
(${\cal N}_\mathrm{fp}$) of zero energy configurations:
\begin{equation}
S\equiv \log({\cal N}_\mathrm{fp}) 
= -\sum_{a \in A } P_{a} ( \{x_l\}_{l\in a}) \ln P_{a} (
\{x_l\}_{l\in a} ) -\sum_{i\in X} ( d_i-1 ) P_i ( x_ i ) \ln P_i
(x_i)\,\,\, .
\end{equation}

\subsection{Exhaustive search}
No known polynomial algorithm is generally able to exhaustively find
all the solutions of a Boolean constraint satisfaction problem like this one.
There is however a number of efficient implementations of exhaustive search
strategies - still exponential in the running times - that allow to
explore problems of reasonable size.

In our case we have mapped the random BN instances onto
\textit{conjunctive normal form} (CNF) formulas.  Such instances are
made of several clauses put in {\sf AND} disjunction, each clause
being made of literals in {\sf OR} conjunction.  This is the natural
form of the well-known satisfiability problem. The mapping is done
writing in the CNF formula all the configurations which violate a
clause in the RBN instance, and negating the literals for the true
variables.  As an example, it is simple to verify that a {\sf AND} node
involving $x_1, x_2$ e $x_3$ as:
$$
x_1 \oplus (x_2 \land x_3)\,\,\, , 
$$
can be written in CNF in the following way:
$$
( x_1 \lor \overline{x_2} \lor \overline{x_3} )
\land
( \overline{x_1} \lor x_2 \lor x_3\nonumber)
\land
(\overline{x_1} \lor  x_2 \lor \overline{x_3})
\land
(\overline{x_1} \lor \overline{x_2} \lor x_3)\,\,\, .
$$ 

Once the problem is cast in this form, we can exploit the vast
number of very well performing algorithms existing for solving
satisfiability instances \cite{SATLIB}.  We are interested now in
\textit{exhaustive search} programs, among which we choose
\texttt{relsat~2.00} \cite{RELSAT}.  This award-winning program can
perform a complete search of the solution space for instances up to
$\sim500$ variables (in our model) in accessible time using a common
pc.

\section{Numerical Results}
\label{sec:num}
In this section we will check the predictive precision of the BP
algorithm measuring the entropy and the magnetization vector.

\subsection{Entropy}
\label{sec:entropy}
In \cite{jstat} it has been pointed out using an
heuristic argument that, for this model of random BNs, the average number of
solutions is always equal to $2^{N-M}$ {\em i.e.} to the number of
all possible external input configurations (note that $N-M$ is exactly
the number of non-regulated sites). A direct numerical check of
Eq.~(\ref{eq:bp}) on single samples shows that the BP predictions are
always in agreement with the above-mentioned heuristic prediction, so
that for any sample, the entropy density $s = S/N = (N-M)/N \log(2) =
(1-\alpha)\log(2)$, independently from the sample realization and the
percentage $X$ of {\sf XOR} functions, apart from terms that vanish when $N \to \infty$. 
\begin{figure}[h]
\includegraphics[width=0.5\columnwidth]{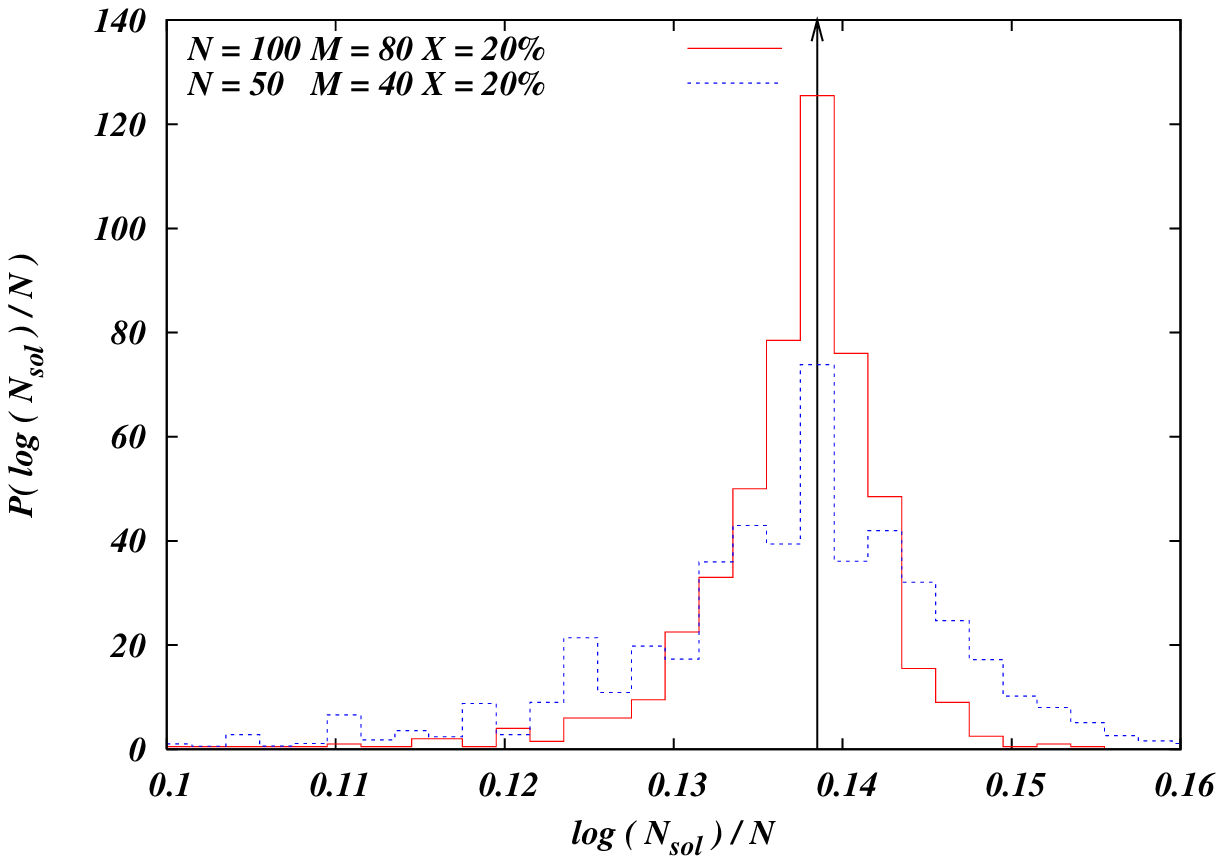}
\includegraphics[width=0.5\columnwidth]{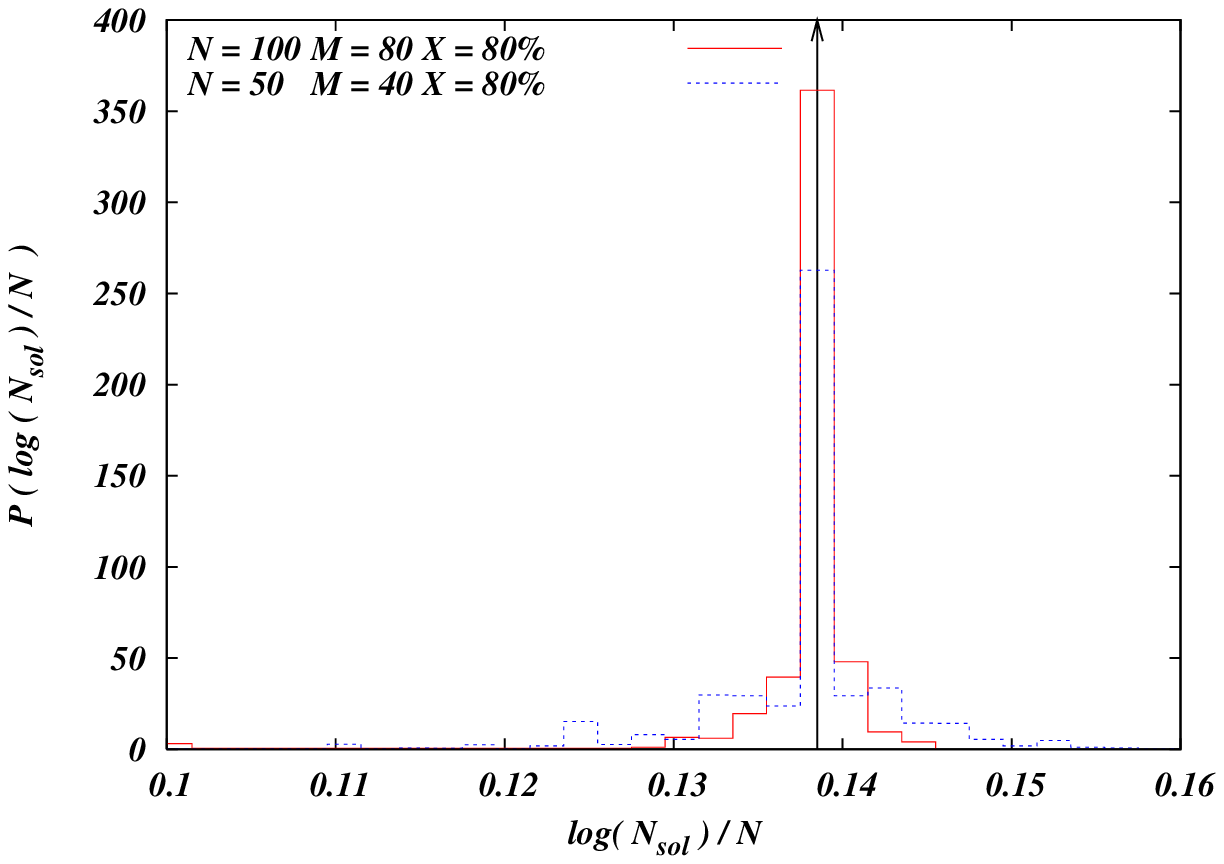}
\includegraphics[width=0.5\columnwidth]{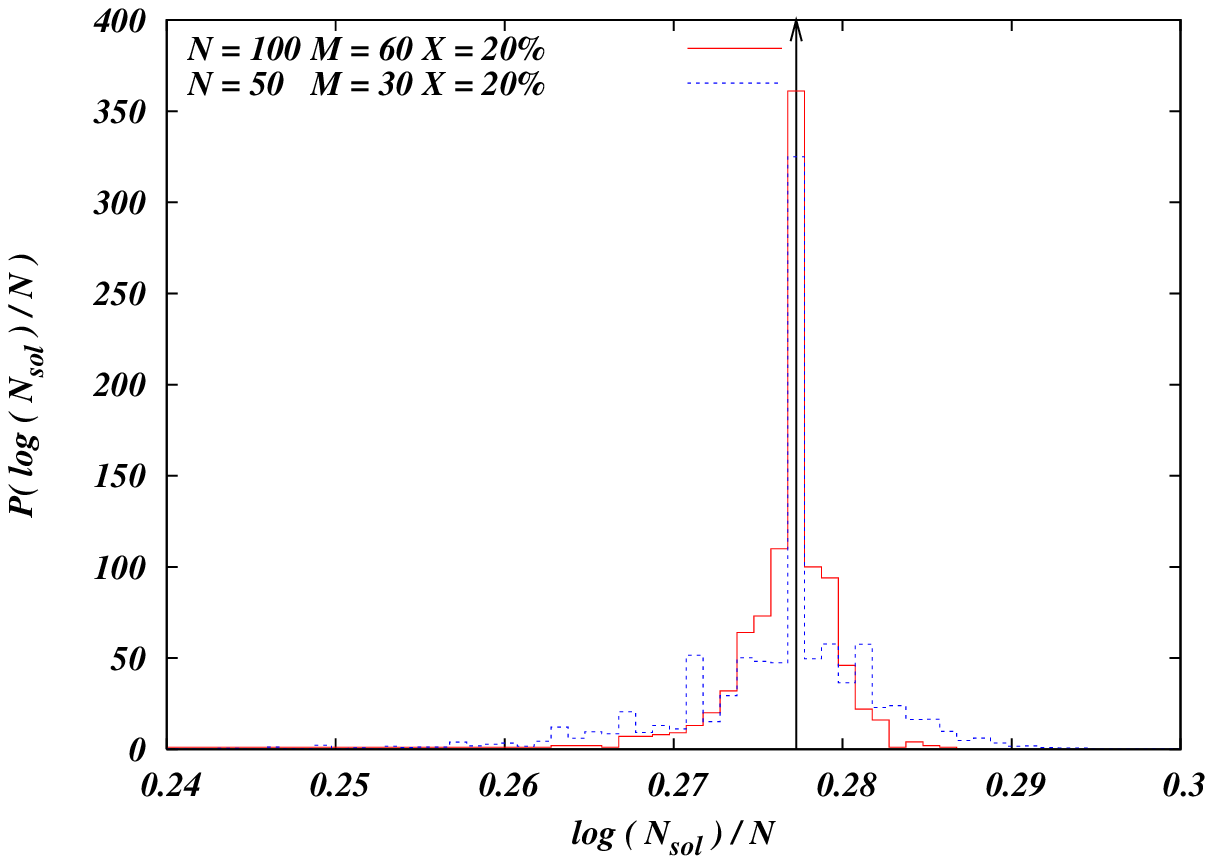}
\includegraphics[width=0.5\columnwidth]{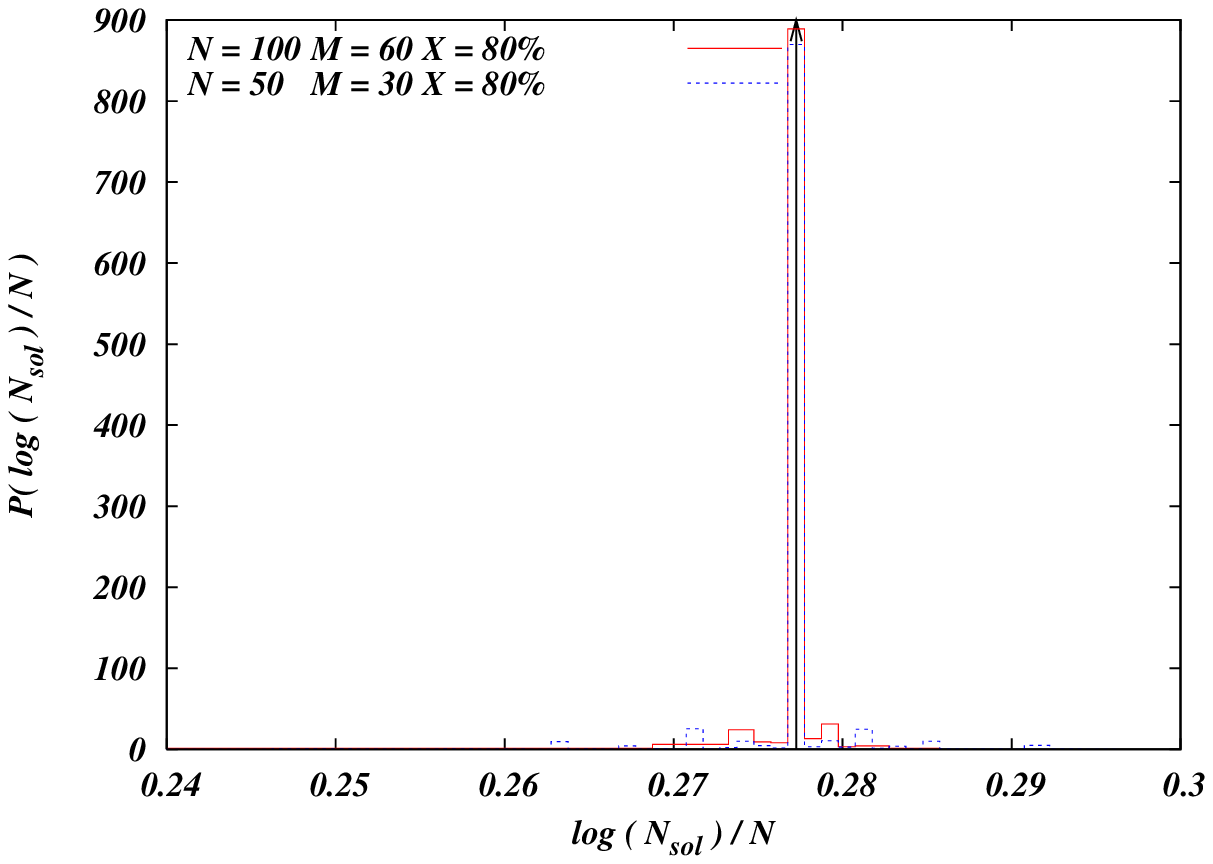}

\caption{Histograms of the exhaustive algorithm estimates measured on
  $10000$ samples of $N=50$ (red boxes) and $N=100$ (blue boxes) for
  four different choices of $\alpha\in\{0.6,0.8\}$ and $X\in\{20\%,
  80\%\}$ expressed as a percentage. The black arrow is the numerical
  estimate using BP, that agrees perfectly with the theoretical value
  $\sigma=(1-\alpha)\ln2$.}
\label{fig:entropy-n50-100}
\end{figure}
In Fig.~(\ref{fig:entropy-n50-100}) we display the frequency
distribution of $\ln \mathcal{N}_{sol}$ for $10000$ samples at
different values of $\alpha$ and $X$. Comparing these histograms
one can observe that
\begin{itemize}
\item The most probable value of $P(s=\ln (\mathcal{N}_{sol})/N)$
depends strongly on $\alpha$ and only mildly on $X$.
\item The distributions at increasing $N$ seem to peak around the
value $s=(1-\alpha)\ln 2$.
\end{itemize}

\subsection{Magnetization}
\begin{figure}[!h]
\includegraphics[width=0.5\columnwidth]{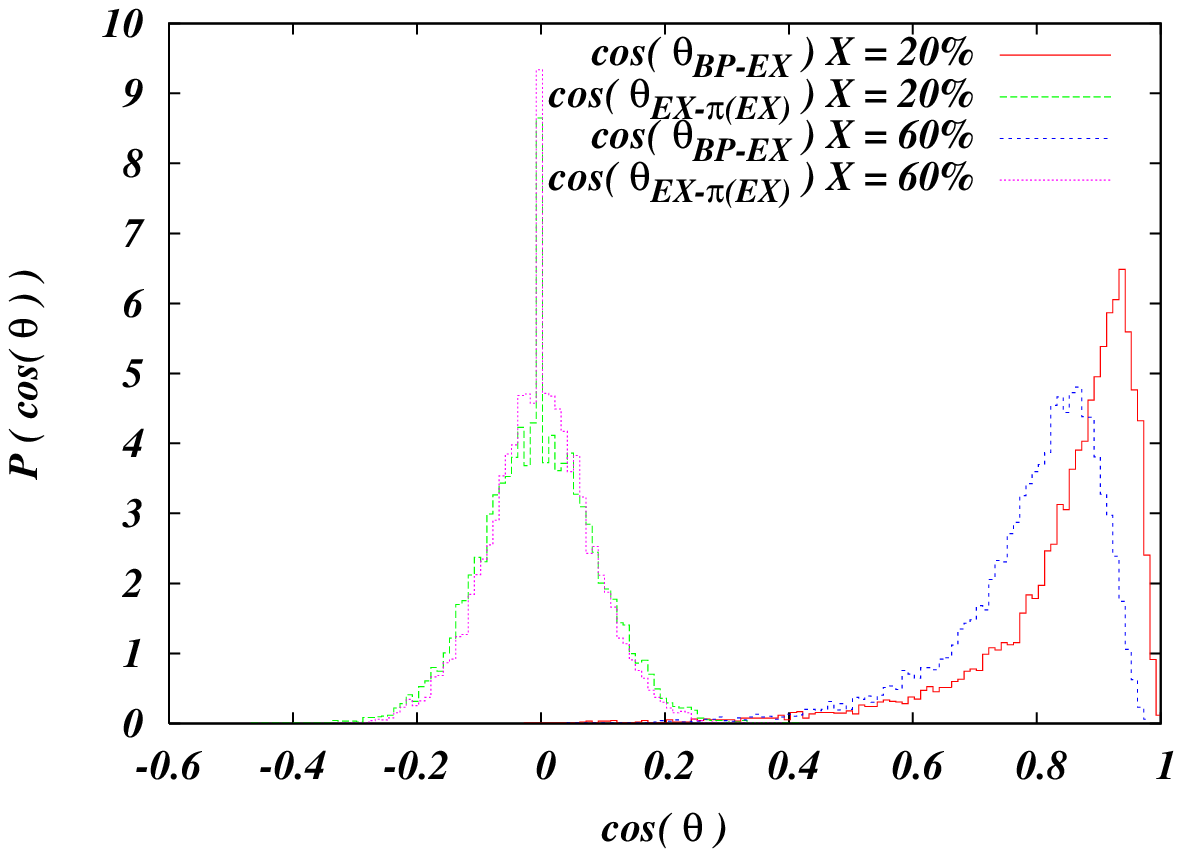}
\includegraphics[width=0.5\columnwidth]{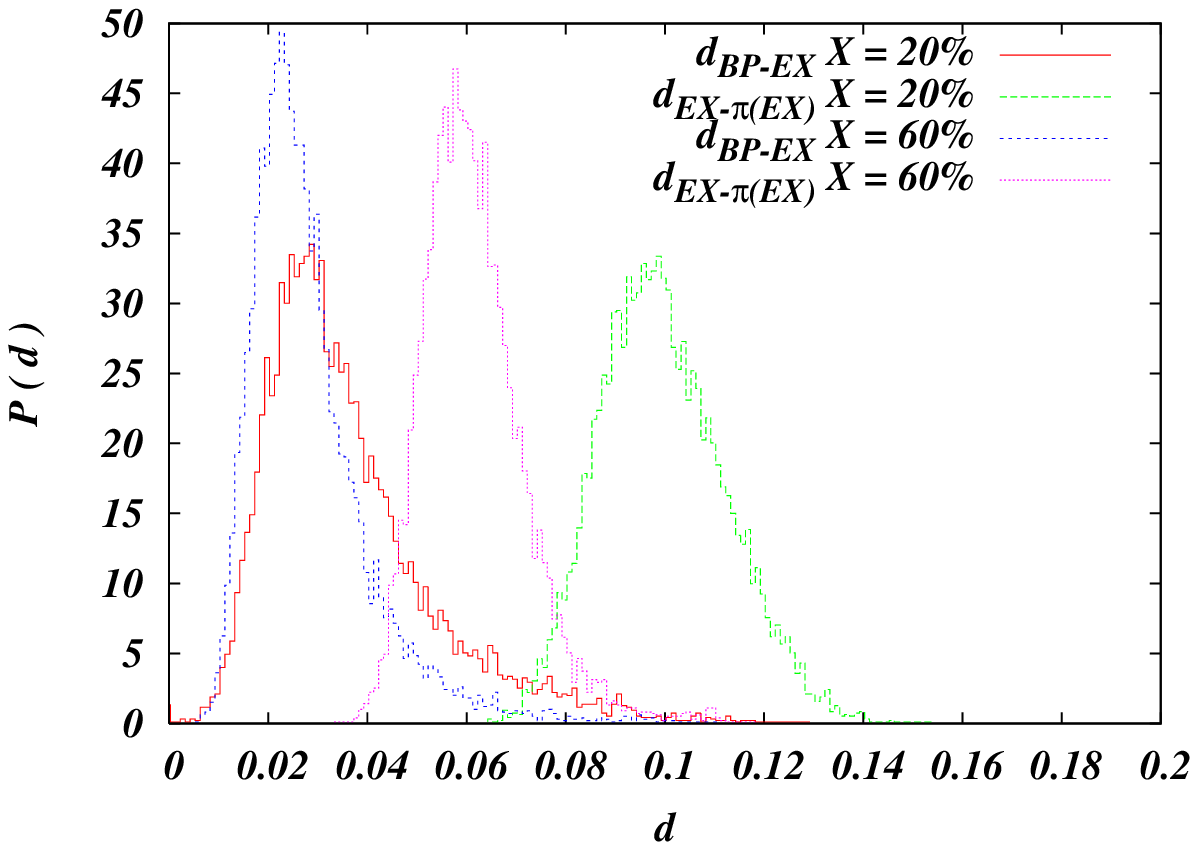}
\caption{ Histogram of $\cos(\theta_{BP-EX})$ compared with the
  reference histogram of $\cos(\theta_{EX-\pi(EX)})$ (left panel). In
  the right panel we display the histogram of both $d_{BP-EX}$ and
  $d_{EX-\pi(EX)}$. Measurement are done at $N=100$ $M=90$ and 
  a percentage of {\sf XOR}
  functions of $X=20\%,60\%$ over an ensemble of $10^4$ samples.}
  \label{fig:mag}
\end{figure}
So far we have analyzed the behavior of the entropy alone.  Although
the entropy is very well predicted by BP, this is not always the case
of the single variables marginal probabilities $P_i ( x_i )$ These
quantities are of extreme importance in any decimation procedure
that is expected to find a specific fixed point via BP.  Necessary
condition for any decimation procedure to be effective is, for each
non pathological sample, a strong positive correlation between
magnetization vectors ${\vec m}^{BP}$ and ${\vec m}^{EX}$,
$N$-dimensional vectors whose $i$-th component represents the
magnetization of each variable $x_i$ defined as
\begin{equation}
m^{BP/EX}_i = P_i^{BP/EX}(1) - P_i^{BP/EX}(0) \,\,\, ,
\end{equation}
where $ P_i^{EX}(x) = P_i^{true}(x) =
\mathcal{N}_{sol}(x_i=x)/\mathcal{N}_{sol}$.  In order to assess this
point, we define two global parameters testing respectively:
\begin{itemize}
\item the probability distribution of the relative angular overlap of
the two magnetization vectors in the $N$ dimensional space, and
\item the probability distribution of relative magnetization euclidean
distances.
\end{itemize}
For each sample, the first overlap parameter is defined
$\cos(\theta_{BP})$, {\em i.e} as the cosine of the angle between the
exhaustive and BP magnetization vectors:
\begin{equation}
\cos(\theta_{BP-EX}) \equiv \frac{\vec{m}^{BP} \cdot
\vec{m}^{EX}}{|{m}^{BP}|\,\,|\vec{m}^{EX}|}\,\, .
\label{overlap1}
\end{equation}
The second is defined as the euclidean distance between non normalized
magnetizations:
\begin{equation}
d_{BP-EX} = \frac{1}{N} \sqrt{\sum_{i=1}^N \left(m^{BP}_i -
m^{EX}_i\right)^2}
\label{overlap2}
\end{equation}
In both cases the predictions have been tested against a null
hypothesis. In the null hypothesis, random magnetization vectors are
extracted in the following way: for each sample ${\vec m}^{\pi(EX)}$
is a random permutation of the components of ${\vec m}^{EX}$, so that
$m^{\pi(EX)}_i\equiv m^{EX}_{\pi(i)}$ where $\pi(i)$ is a random
permutation of the ordered set $\{1,\dots,N\}$. The quantities
$\cos(\theta_{EX-\pi(EX)})$ and $d_{EX-\pi(EX)}$ are then calculated
substituting ${\vec m}^{\pi(EX)}$ to ${\vec m}^{BP}$ in
eqs.(\ref{overlap1}) and (\ref{overlap2}). Distributions of the
overlaps over the sample populations are plotted in \fref{fig:mag}.
The results show a strong correlation between the true and the
predicted magnetization vector distributions.

\subsection{Solutions overlap}

It has been indicated in \cite{letter,jstat} that in this model  the
entropy is analytic in all the phase diagram, while the organization of the fixed points
undergoes a sudden reorganization at some $\alpha_d (X)$:  
\begin{itemize}
\item At $\alpha<\alpha_d(X)$ all solutions are in a single cluster, {\em i.e.}  any pair of solution
is connected by a path via other solutions, where in each step only a finite 
number of variables can be changed.
\item At $\alpha>\alpha_d(X)$ The space of solutions spontaneously breaks into an exponential number of macroscopically separated clusters of fixed points. Their number, or more precisely its normalized logarithm, is called complexity. It is a first-order phase transition.
\end{itemize}

\begin{figure}[ht]
\includegraphics[width=0.33\columnwidth]{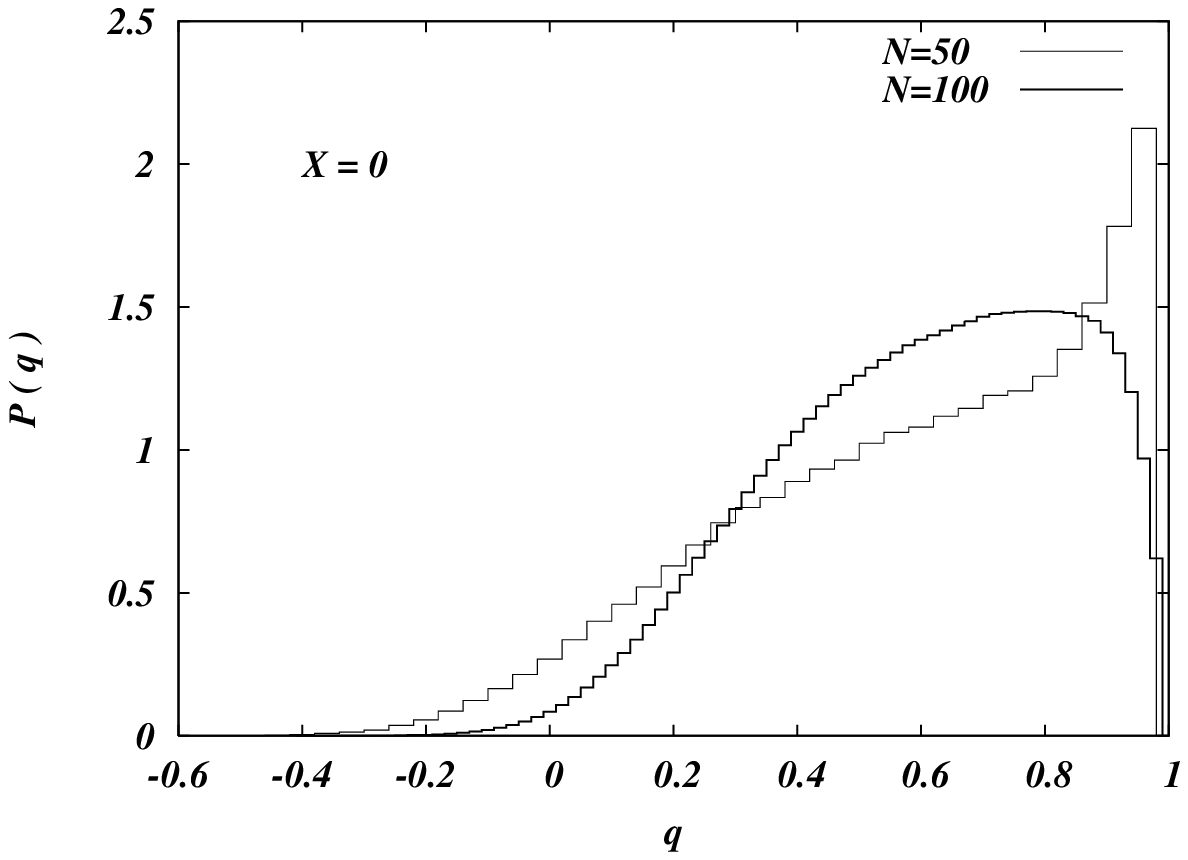}
\includegraphics[width=0.33\columnwidth]{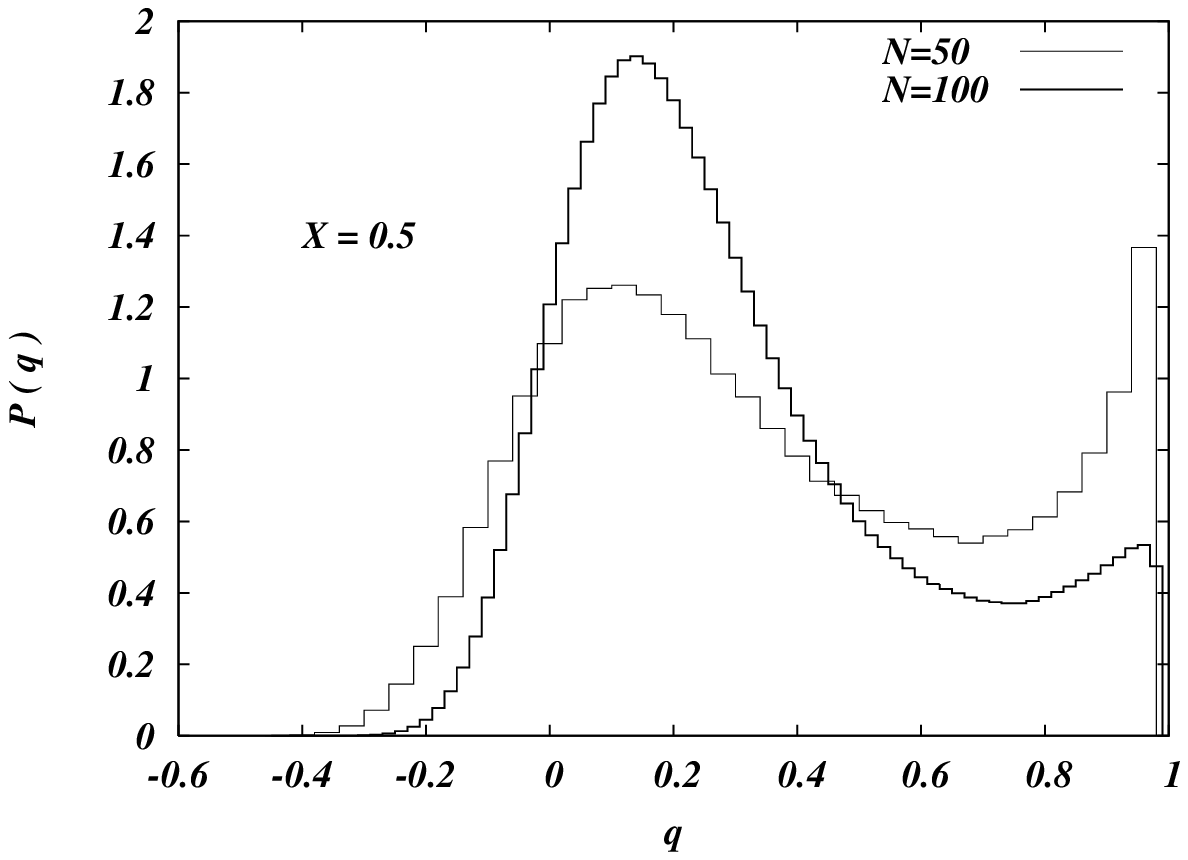}
\includegraphics[width=0.33\columnwidth]{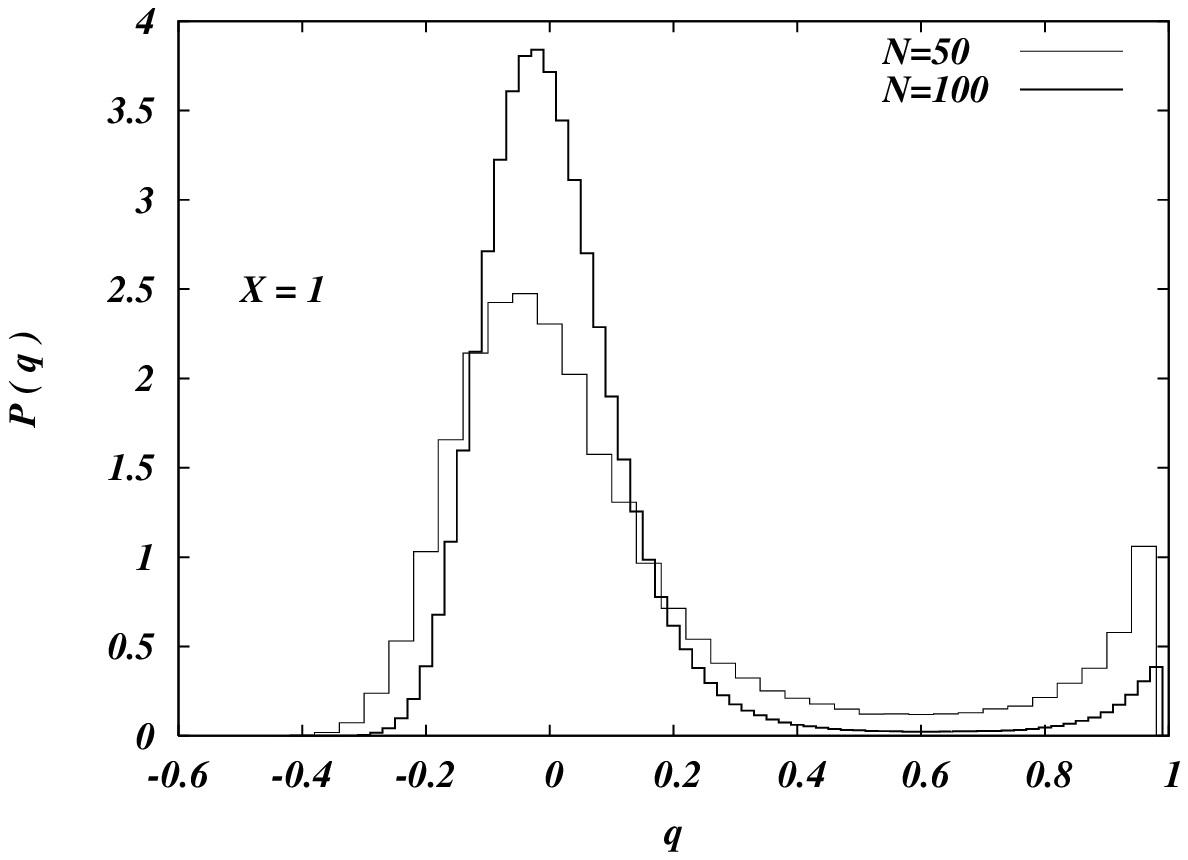}
\caption{Distribution of the overlaps for 10000 samples at
$\alpha=0.93$. $X=0$ (left panel), $X=0.5$ (central panel), $X=1$
(right panel), for $N=50$ (thin line) and $N=100$ (thick line).}
  \label{fig:overlap}
\end{figure}

This behavior is characterized by the appearance
of a non-trivial structure of the space of solutions. In other words
the fixed points, rather than being uniformly scattered over the
$N$-dimensional hypercube, start to organize themselves in clusters,
with a well defined \textit{intra-cluster} and \textit{inter-cluster}
overlap.  Nevertheless, by means of the exhaustive enumeration
technique introduced above, one can easily write down all the
solutions for a given sample and hence calculate all the ${\cal
N}_\mathrm{sol}({\cal N}_\mathrm{sol}-1)/2$ overlaps, defined in a
standard way as
$q^{ab}=\frac{1}{N}\sum_{i=1}^N\sigma^a_i\sigma^b_i$. Note that we are
using spin variables here $\sigma_i=-1+2{x_i}$, and $a,b$ indicates
two distinct solutions.  The distribution of the overlaps for two sizes
and three different choices of $X$ are shown in \fref{fig:overlap}. We
choose a value far apart form the clustering transition line (no {\sf XOR}
functions), a value in the vicinity of the transition line ($50\%$ of
{\sf XOR} functions) and a value deep inside the clustered phase
($100\%$ of {\sf XOR} functions).

It is clear that as one moves from the unclustered to the clustered phase
(which at fixed $\alpha$ is equivalent to increasing $X$), the 
distribution changes from a broad shape to a
two-peaked one. However, this two-peaked pattern is just a finite size
effect, as it seems to be indicated by the reduction of the weight of the right peak
of the $P(q)$ from $N=50$ to $N=100$ shown in the left panel of
\fref{fig:overlap}. As the number of clusters is exponential in the
system size, the probability of extracting two random solutions from
the same cluster is negligibly small even for moderate sizes.  

This phenomenon can be understood by the following qualitative argument: let
us suppose that the number of clusters as well as the size of each single cluster
are exponential in the system size. Let us further suppose that
the distribution of cluster sizes is strongly peaked around a single value,
${\cal N}_{\mathrm{sol-in-cl}}$ (less stringent conditions can be found,
together with pathological cases where those conditions do not hold,
but a complete treatment would go beyond the scope of present work).
In a completely clustered phase the weight contribution of the
intra-cluster overlap will then be proportional to ${\cal N}_{\mathrm{cl}} {\cal
N}^2_{\mathrm{sol-in-cl}} $ where ${\cal N}_{\mathrm{cl}}$ is the
number of clusters. On the other hand, 
the contribution to the inter-cluster overlap is proportional to ${\cal N}^2_{\mathrm{cl}} {\cal
N}^2_{\mathrm{sol-in-cl}}$ ({\em i.e.} all couple of solutions except
those in the same clusters). 
If ${\cal N}_{\mathrm{cl}}$ is exponential in $N$, the
intra-cluster contribution will be negligible with respect to the
inter-cluster one in the thermodynamic limit.

\section{Some considerations on the validity of the annealed approximation}
\label{sec:theo}
\label{NSOL_DIST}

In this section we will show that the logarithm of the number of solutions ({\em i.e.} the entropy)
of our model of BNs computed within the annealed approximation scheme 
agrees with the numerical estimate for the entropy we performed 
by exhaustive enumeration and by 
the numerical solution of the BP equations on single sample. 

This feature allows us to
make some conjectures, and check whether the annealed 
approximation is exact in the $N \to \infty$
limit. We will show that the variance of $P( \mathcal{N}_{sol})$ as
well as all higher moments are proportional to $2^{(1-\alpha) N}$ in
the large $N$ limit, with a proportionality constant depending only on
$\alpha$ and $X$.  This implies that a non zero contribution to the
entropy is given at the leading order only by the external regulating
variables, in accordance with results obtained in \cite{letter,jstat}.
However, we will also show how, in the general case of $X < 1$, the
proportionality constants are different from one and greater than it,
not ensuring the exactness of the annealed result, albeit replica
results given in \cite{letter,jstat} are a strong hint in that
direction.

Moreover, we will see that in the thermodynamic limit:
\begin{equation}
\frac{\langle \mathcal{N}^2_{sol}
\rangle}{\langle \mathcal{N}_{sol}\rangle^2} \equiv C(\alpha, X)
\label{diverge}
\end{equation}
where $C(\alpha, X)$ is a constant independent of $N$.
Only in the case of pure {\sf XOR} classes ($X=1$), it is possible to
show that the proportionality constant is exactly $1$ the large $N$
limit for any order moment and any value of $\alpha$, implying that
the annealed approximation is exact. 

   Let's consider a distribution of positive random variables $Z_i$. As                                            
   we are interested in the relation between the quenched entropy and the                                          
   annealed approximation, we want to investigate the relation between                                             
   $\langle \ln Z\rangle$ and $\ln \langle Z \rangle$, where                                                       
   $\langle\phantom{Z}\rangle$ identifies here the average over the                                                
   distribution of $Z_i$. We can write:                                                                            
   \begin{equation}                                                                                                
   \langle \ln Z \rangle = \ln \langle Z \rangle +\langle \ln \frac{Z}{\langle Z \rangle} \rangle =                
   \ln \langle Z \rangle +                                                                                         
   \sum_{i=2}^{\infty}\frac{(-1)^{i+1}}{i}\langle\Big(\frac{Z-\langle Z \rangle}{\langle Z \rangle}\Big)^i\rangle  
   \end{equation}                                                                                                  
   where we have expressed the quenched entropy in term of the annealed                                            
   one plus a series of the moments of the distribution. In our case we                                            
   take $Z = \mathcal{N}_{sol}$. In cases where the sum of the series of                                           
   moments is finite or for any $X$ and $\alpha$ diverges as a function                                            
   $f(N)$ of the size of the graph such that $f(N)/N \to 0$, then the                                              
   quenched and the annealed entropies coincide. We will show that this                                            
   is the case at least for the second order moment. It is important to                                          
   state, however, that the previous series expansion is valid only if
   $0 < Z/\langle Z \rangle \le 2$, and that averaging the resulting series                                             
   term by term is possible only if the sum can be taken out of the integral                                       
   which defines the average.                                                                                      

These conditions are not always met in our model, and in particular we
expect the existence of a certain range of $X$ and $\alpha$ values
beyond which the conditions are not necessarily satisfied. In particular 
we will also show that in the thermodynamic limit
\begin{equation}
\lim_{X  \to 0}  \lim_{\alpha \to 1}C(\alpha, X) = \infty
\end{equation}
However this will not necessarily undermine the validity of the annealed calculation.

Under the hypothesis of equation (\ref{diverge}), and  thanks to 
a straightforward implementation of the Tschebichev inequality,  one can easily show that, (see \ref{app:cheby}) 
\begin{equation}
\mbox{Pr}(\mathcal{N}_{sol} > \langle \mathcal{N}_{sol}\rangle ) \leq 2^{-N\gamma} C(\alpha, X) 
\end{equation}
where $\gamma$ is a positive constant. This implies that, in the thermodynamic limit, the probability distribution of the number of solutions has support in the interval $(0, \langle\mathcal{N}_{sol}\rangle)$. Unfortunately in order to take under control also the left tail of the distribution one should compute moments of the type $\langle \mathcal{N}_{sol}^n \rangle $ with $0<n<1$, which is, as indicated in \ref{app:high_mom}, a rather complicated task.

\subsection{General calculation of $\langle \mathcal{N}_{sol} \rangle$}

Given a probability distribution of classes of Boolean functions $\pi
(f)$, drawn independently, where $f$ is a generic Boolean function of
$K=2$ inputs, and given a value of $\alpha \in [0,1]$ one can write
\begin{equation} 
\langle \mathcal{N}_{sol} \rangle = \sum_{\vec{x}} \langle 
\prod_{m=1}^{M \sim \alpha N} \langle \delta(1; x_{0,m} f(x_{1,m},x_{2,m})) \rangle_{\pi(f)}
\rangle_{\cal{G}}
\label{eeq1}
\end{equation}
where the external average is over the graph ensemble, while the
internal one is over $\pi(f)$.  Note that for the {\sf XOR} class
$f(x_1,x_2) = \epsilon x_1 x_2$, while for the {\sf AND-OR} class
$f(x_1,x_2) = \epsilon_0 /(2 (\epsilon_1 x_1 \epsilon_2 x_2 +
\epsilon_1 x_1 + \epsilon_2 x_2 -1))$, with $\epsilon_i \in \{ \pm 1
\}$ with a chosen probability and $X \in [0,1]$. The variables $\{ x_i
\} \in \{ \pm 1 \}^N$. Averaging over the values of $\{ \epsilon_i
\}_{i=0,1,2}$ and $X$ is therefore equivalent of averaging over the
$\pi(f)$.  In particular, in the case of flat classes distribution,
which is the object of the present work, one has $prob( \epsilon ) =
(\delta(\epsilon - 1) + \delta(\epsilon + 1) )/2$. Therefore:
\begin{equation} 
\langle \mathcal{N}_{sol} \rangle = \sum_{\vec{x}} \langle 
\prod_{m=1}^{M \sim \alpha N} \langle \delta(1; x_{0,m} f(x_{1,m},x_{2,m})) \rangle_{\{\epsilon_i\},X}
\rangle_{\cal{G}}
\label{eeq1b}
\end{equation}
Let us now define $N_o = \alpha N$ regulated variables and $N_i =
(1-\alpha) N$ external inputs (including the $(1-\alpha) e^{-2\alpha}$
isolated nodes). Assuming the input variables are extracted randomly
and independently on each of the $N_o$ clauses; one can write
\begin{eqnarray}
\langle \mathcal{N}_{sol} \rangle &=& \sum_{N^+_o,N^+_i=0}^{N_o,N_i}
{N_o \choose N^+_o} { N_i \choose N^+_i} \nonumber \\ 
& & \left[
{\cal P}(++|+)g(+++) +  {\cal P}(+-|+)g(+-+) + \right. \nonumber \\
& & \left. {\cal P}(-+|+)g(-++) + {\cal P}(--|+)g(--+) 
\right]^{N^+_o} \nonumber \\   
& & \left[
{\cal P}(++|-)g(++-) +  {\cal P}(+-|-)g(+--) + \right.\nonumber \\
& & \left.  {\cal P}(-+|-)g(-+-) + {\cal P}(--|-)g(---) 
\right]^{N_o - N^+_o} 
\label{eeq2}
\end{eqnarray}
where ${\cal P} (\rho \sigma|\tau)$ is the probability of drawing two
inputs of sign $\rho$ and $\sigma$ given the fact that one is looking
at a clause with output variable sign $\tau$, and $g(\rho \sigma
\tau)$ is the value of the function node
$\delta(1;\tau f(\rho,\sigma))$
times the probability of extracting a certain
Boolean function type $f$.  In the case of uniform $\pi (f)$, ${\cal
P} (\rho \sigma|\tau)g(\rho \sigma \tau)$ is trivially $1/8$ $\forall$
signs triplet, leading to $\langle \mathcal{N}_{sol} \rangle =
2^{(1-\alpha) N}$ identically.

For a different distribution this is in general {\it not } the case.
As a title of example, in the case of a mixture of pure {\sf XOR} and
{\sf AND} functions, without any literal negation, $\pi (f) =
X\delta(f;f_{\oplus}) + (1-X)\delta(f;f_{\wedge})$ and eq.(\ref{eeq2})
reads
\begin{eqnarray}
\langle \mathcal{N}_{sol} \rangle &=& \sum_{N^+_o,N^+_i=0}^{N_o,N_i}
{N_o \choose N^+_o} { N_i \choose N^+_i} \cdot \nonumber \\ 
& & \left[ 
\frac{2 X (N-N^+)(N^+-1)}{(N-1)(N-2)} + (1-X)\frac{(N^+ - 1)(N^+ - 2)}{(N-1)(N-2)}
\right]^{N^+_o} \cdot \nonumber \\   
& & \left[
X\frac{N^+(N^+-1)}{(N-1)(N-2)} + \frac{(N-N^+-1)(N-N^+-2)}{(N-1)(N-2)} \right. \nonumber \\
& & + \left. 
2 (1-X)\frac{N^+(N-N^+-1)}{(N-1)(N-2)}
\right]^{N_o - N^+_o} 
\label{eeq2b}
\end{eqnarray}
Leading order terms can be computed following a calculation identical
to the one shown below for the second moment, and will be omitted
here.

\subsection{General calculation of $\langle \mathcal{N}^2_{sol} \rangle$}  

\begin{figure}[ht] 
   \centering
   \includegraphics[width=12cm, angle=0]{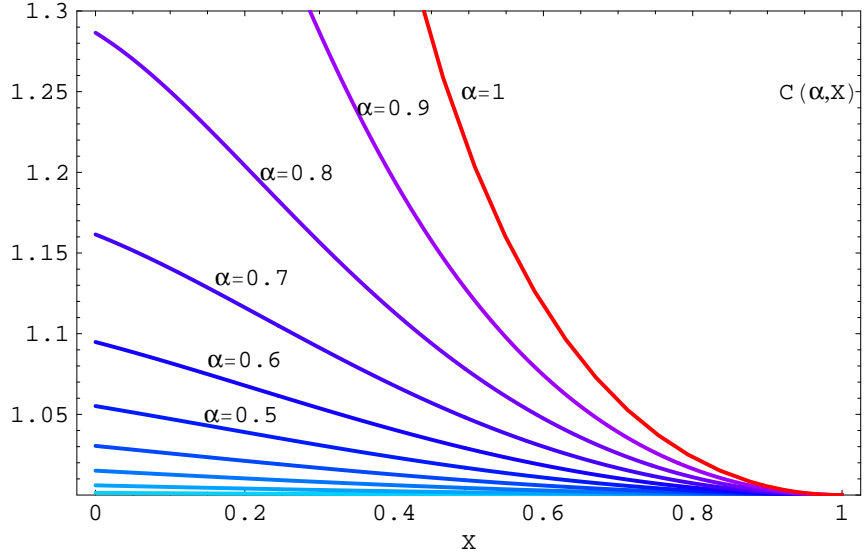}
   \caption{Plot of $C(\alpha,X)$ for the second moment of the
distribution of the number of solutions in the large $N$ limit. The
red uppermost line is the function $e^{(x-1)}/x$, diverging for $x \to
0$}
   \label{figura-m1}
\end{figure}

For the second moment:
\begin{equation} 
\langle \mathcal{N}^2_{sol} \rangle = \sum_{\vec{x}, \vec{y}} \langle 
\prod_{m=1}^{M \sim \alpha N} \langle \delta(1; x_{0,m} f(x_{1,m},x_{2,m}))
\delta(1; y_{0,m} f(y_{1,m},y_{2,m}))
\rangle_{\{\epsilon_i\},X}
\rangle_{\cal{G}}
\label{eeq3}
\end{equation}
Averaging uniformly over the function types one obtains
\begin{equation} 
\langle \mathcal{N}^2_{sol} \rangle = \sum_{\vec{x}} \langle 
\prod_{m=1}^{M \sim \alpha N} \mathcal{G}^{(2)}(X;x_{0,m},x_{1,m},x_{2,m})
\rangle_{\cal{G}}
\label{eeq4}
\end{equation}
with
\begin{equation} 
\mathcal{G}^{(2)}(X;x_0,x_1,x_2) = \frac{X}{2}(1 + x_0 x_1 x_2) + 
\frac{1-X}{2}((1 + x_1 + x_2 + x_1 x_2)\frac{x_0}{4} + 1)
\label{eeq5}
\end{equation}
Averages over different function types distributions are also
possible, but beyond the scope of this work.  Along the same line of
arguments of the previous section, averaging over the Poissonian graph
structure, one obtains
\begin{eqnarray}
\langle \mathcal{N}^2_{sol} \rangle &=& \sum_{N^+_o,N^+_i=0}^{N_o,N_i}
{N_o \choose N^+_o} { N_i \choose N^+_i} \nonumber \\ 
& & \left[
{\cal P}(++|+)g(X;+++) +  {\cal P}(+-|+)g(X;+-+) + \right. \nonumber \\
& & \left. {\cal P}(-+|+)g(X;-++) + {\cal P}(--|+)g(X;--+) 
\right]^{N^+_o} \nonumber \\   
& & \left[
{\cal P}(++|-)g(X;++-) +  {\cal P}(+-|-)g(X;+--) + \right.\nonumber \\
& & \left.  {\cal P}(-+|-)g(X;-+-) + {\cal P}(--|-)g(X;---) 
\right]^{N_o - N^+_o} 
\label{eeq6}
\end{eqnarray}
where $g(X;\rho\sigma\tau)$ is now the value of $\mathcal{G}^{(2)}$
given $X$ and the sign of the inputs. Equations (\ref{eeq2}) and
(\ref{eeq6}) have an identical structure. This observation can be
generalized and holds for any higher order momentum, changing the
structure of the functions $g$.  The details of the computation of the
second moment are reported in \ref{sec:app}. At the leading order it
turns out that:
\begin{equation}
\langle \mathcal{N}^2_{sol} \rangle \equiv I^{(2)}(\alpha,X) = 2^{2 (1-\alpha) N} C(\alpha,X)
\label{eqsaddle1b}
\end{equation}
The value of the multiplicative constant $C(\alpha,X)$ as a function
of $X$ for increasing values of $\alpha$ is plotted in
fig.(\ref{figura-m1}).
Note that, unless in the pure {\sf XOR} case, $C(\alpha,X) \ne 1$. This
means that the analysis of the second order momentum is not enough to
assess the theoretical validity of the annealed calculation of the
entropy in the sense that the convergence of the logarithmic
correction series is not ensured a priori. Moreover, whenever
$C(\alpha,X) > 1$, $\sigma^2_{{\cal N}_{sol}} = \langle
\mathcal{N}^2_{sol} \rangle - \langle \mathcal{N}_{sol} \rangle^2
\propto 2^{N(1-\alpha)}$.

\subsubsection{The special case of $\alpha = 1$}
At $\alpha = 1$ several simplifications in the computation of the second moment
hold. Both the entropic contribution due to the external regulators, and  
the Kullback-Leibler terms are identically  zero at
the saddle point, as one can check from the saddle
point equation for $\langle \mathcal{N}^2_{sol} \rangle$ presented in
\ref{sec:app}.  
It turns out that the contribution to the saddle point includes
also the term corresponding to the non-zero boundary term of integration.   
For the second moment in
particular, one has to take into account the contribution of all terms
around $N_o^+ = N_o$.  

\begin{figure}[ht] 
   \centering \includegraphics[width=12cm, angle=-90]{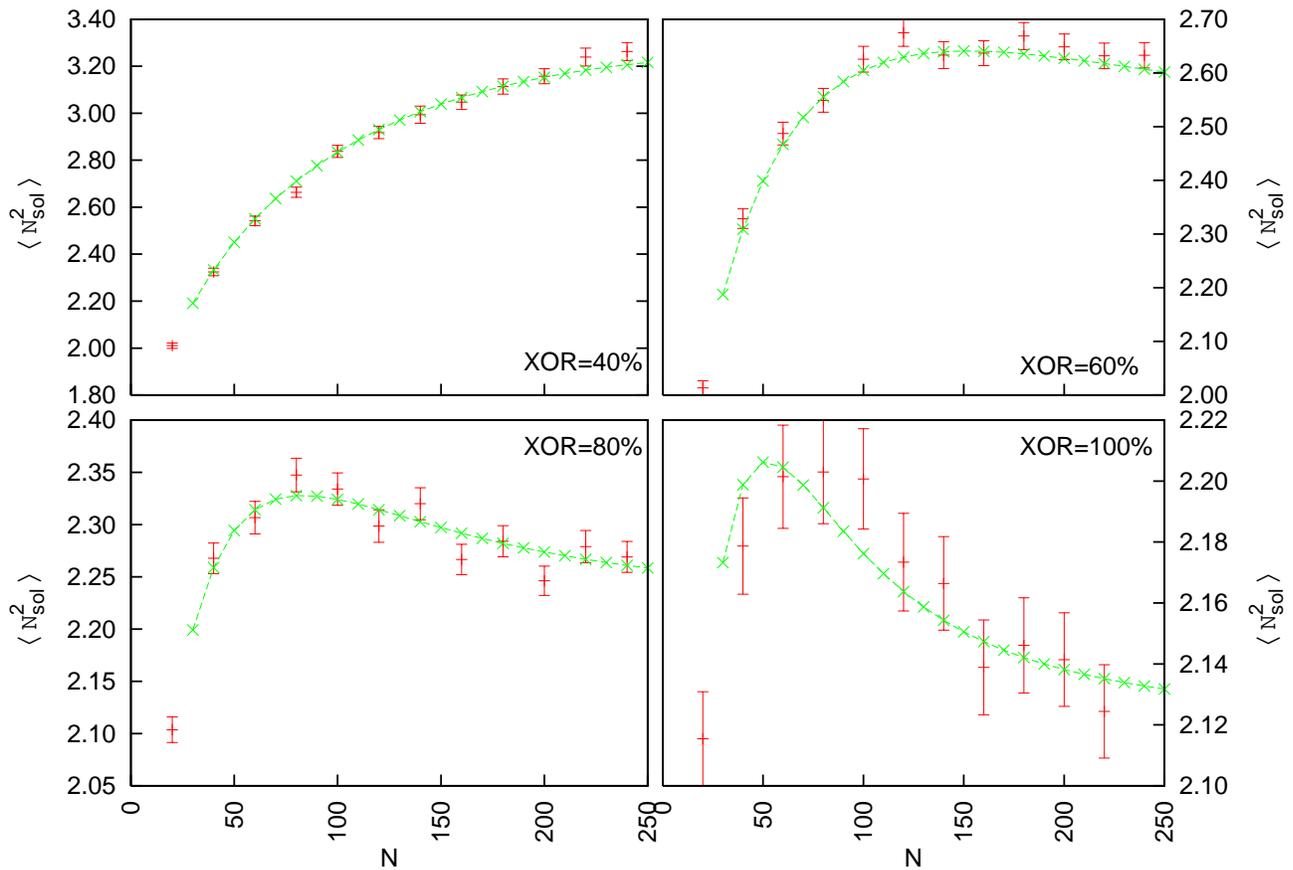}
   \caption{Second moment of the number of solutions
            distribution. Bars are computed by exhaustively counting
            the solutions for 100000 samples at $\alpha=1$, crosses by evaluating
            the analytical formula with Mathematica.}
   \label{N_SQUARE}
\end{figure}

Going back to eq.(\ref{eeq6}) and taking
explicitly into account those terms, one obtains in the large $N$
limit
\begin{equation}
\label{eq:n2}
\langle \mathcal{N}^2_{sol} \rangle \to I^{(2)}(1,X) + \sum_{t=0}^{\infty} \frac{e^{-t(X+1)}(t-1)^t (X+1)^t}{t!}
\end{equation}
where $I^{(2)}(1,X) = e^{(x-1)}/x$.  The values of both contributions
diverge for $X \to 0$.  Analytic estimates suggest that
the divergence goes as $\sqrt{N}$. In fig.~\ref{N_SQUARE} we display
the numerical estimate of $\langle \mathcal{N}^2_{sol} \rangle$
obtained via exhaustive enumeration and the analytic estimate
presented in eq.~(\ref{eeq6}).
The agreement is good, as it should be, since the computation of the second
moment is exact, also for finite $N$. 
\section{Conclusions}
In this work we have investigated several aspects of the fixed points
solutions set of a particular class of {\it finite size} diluted
random Boolean networks with $K=2$ inputs per function. A search
algorithm based on a state-of-the-art satisfiability solver was used
to exhaustively enumerate fixed points up to moderate system
sizes. For fixed system size ensembles the average number of solutions
were plotted together with average fluctuations and with two
additionally ad hoc defined order parameters indicating average
distance between solutions within the fixed points set. A throughout
comparison of the exact results with those given by the heuristic
Belief Propagation algorithm was done in order to assess the
performance of BP for small size samples whose network structure
significantly deviates from tree-like. BP was shown to perform
significantly well in the prediction of the correct number of
solutions even in small sizes cases. BP seems to loose its predicting
power in the calculation of single Boolean variables marginals at the
fixed points; still it was shown to retain a significant correlation
with exact results in the global spatial arrangements of the
solutions. Furthermore, the analytical results closing the paper,
together with their agreement with exact enumeration results, give a
strong hint on the exactness of the annealed calculation of the
entropy as well as the high order moments of the probability
distribution of the number of solutions.

\section{Acknowledgments}
We are happy to acknowledge Martin Weigt for his suggestions  about the computation
of the first and second moment. We want also to thank Matteo Marsili,  and Riccardo Zecchina,
for many interesting discussions.

\appendix
\section{The computation of the moments of $P(\mathcal{N}_{sol})$}
\label{sec:app}
We first present the details of the computations of the second moment
and we will then give some hints on the structure of the generic
$n^{th}$ moment.
\subsection{ The second order momentum }
With respect to $\langle \mathcal{N}^2_{sol}\rangle$, if we keep only
the leading order terms in eq.(\ref{eeq6}), the sums can be
approximated by the following integral, where the exponent is given by
the entropy contribution of the external variables plus a
Kullback-Leibler distance term vanishing at the saddle point:
\begin{eqnarray}
\langle \mathcal{N}^2_{sol} \rangle = I^{(2)}(\alpha,X) &=&
2^{(1-\alpha)N}\int_0^1\int_0^1 db_o^+db_i^+ K(\alpha,X,b_o^+,b_i^+)
e^{N {\cal F}[\alpha,X,b_o^+,b_i^+]} \nonumber \\
K(\alpha,X,b_o^+,b_i^+) &=& \frac{\sqrt{\alpha
(1-\alpha)}}{\sqrt{b_o^+ b_i^+(1-b_o^+)(1-b_i^+)}} e^{\alpha
B(\alpha,X,b_o^+,b_i^+) } \nonumber \\ {\cal
F}[\alpha,X,b_o^+,b_i^+] &=& (1-\alpha)H(b_i^+) + \alpha D_{KL}
(b_o^+|G(\alpha,X,b_o^+,b_i^+)) \nonumber \\ B(\alpha,X,b_o^+,b_i^+)
&=& 3 - \frac{3 b^+ b_o^+ + (1-X)(1-b^+)b_o^+ +
(1+X)(1-b^+)b_o^+/2}{(b^+)^2+(1-X)(1-b^+)b_o^+ + (1+X)(1-b^+)^2/2}
\nonumber \\ &-& \frac{(1+X)(b^+(1-b_o^+) +
3(1-X)(1-b^+)(1-b_o^+)/2}{1 - (b^+)^2+(1-X)(1-b^+)b_o^+ -
(1+X)(1-b^+)^2/2} \nonumber \\ H(x) &=& -x \log(x) - (1-x) \log(1-x)
\nonumber \\ D_{KL} (x|y) &=& x\log\left( \frac{y}{x} \right) +
(1-x)\log\left( \frac{1-y}{1-x}\right) \nonumber \\
G(\alpha,X,b_o^+,b_i^+) &=& (b^+)^2 + (1-X)b^+(1-b^+) +
\frac{1+X}{2}(1-b^+)^2 \nonumber \\ b^+(\alpha,b_o^+,b_i^+) &=& \alpha
b_o^+ + (1-\alpha)b_i^+ \nonumber
\label{eeq8}
\end{eqnarray}
where $b_o^+ = N_o^+/N_o$ and $b_i^+ = N_i^+/N_o$.  The value of the
Integral $I^{(2)}(\alpha,X)$ can be calculated at the saddle point
\begin{eqnarray}
{\tilde b}_i^+ &=& \frac{1}{2} \\ {\tilde b}_o^+ &=&
G(\alpha,X,{\tilde b}_o^+,{\tilde b}_i^+) \\ &=& \frac{2 -
\alpha(1-\alpha) + \alpha X (1 + 3\alpha) - 2 \sqrt{1-\alpha(1-X) +
\alpha^2 X(X-1)}}{2 \alpha^2 (1 + 3 X)} \nonumber
\label{eqsaddle1}
\end{eqnarray}
Finite $N$ corrections can be in principle computed extending the
calculation of $K$ to higher orders in ${\cal O}(1/N)$, and performing
an asymptotic expansion around the saddle point. Whenever $\alpha < 1$
and $X > 0$ it can be seen that the condition (\ref{eqsaddle1}) finds
the only maximum ${\cal F}$, which lies within the integration
interval.
\subsection{An upper bound on the number of solutions}
\label{app:cheby}
In this subsection we will show that in the thermodynamic limit the support of $P(\mathcal{N}_{sol})$ is contained in the interval
$(0 , \langle \mathcal{N}_{sol} \rangle )$, using, as we have already shown in the previous section,  that:
\begin{eqnarray}
\langle \mathcal{N}_{sol} \rangle &=& 2^{(1-\alpha)N} \nonumber \\
\langle \mathcal{N}^2_{sol} \rangle &=& 2^{2(1-\alpha)N} C(\alpha, X) 
\label{eq:2mom}
\end{eqnarray}
where $C(\alpha,X)$ is a constant independent from $N$. The one-tailed Chebyshev inequality states that:
\begin{equation}
\mbox{Pr} \left( \mathcal{N}_{sol} >  \langle \mathcal{N}_{sol} \rangle \right) \leq \frac{ \langle  \mathcal{N}^2_{sol} \rangle }
{  (\mathcal{N}_{sol} - \langle  \mathcal{N}_{sol} \rangle)^2 }\,\,\,\,.
\label{eq:cheby}
\end{equation}
Under the condition $ \mathcal{N}_{sol} >  \langle \mathcal{N}_{sol} \rangle $ we can express $\mathcal{N}_{sol} = 2^{N\Sigma'}$ where $\Sigma' > (1-\alpha)$. Inserting eqs.~(\ref{eq:2mom}) into eq.~(\ref{eq:cheby}) we get:
\begin{eqnarray}
\mbox{Pr} \left( \mathcal{N}_{sol} >  \langle \mathcal{N}_{sol} \rangle \right) &\leq& 
\frac{2^{2(1-\alpha)N} C(\alpha, X)}{2^{2N\Sigma'} (1-2^{N(1-\alpha - \Sigma')} )} \leq  2^{2N (1-\alpha - \Sigma')} C(\alpha, X)\nonumber \\
&=& 2^{-N \gamma} C(\alpha,X)\,\,\,\,,  
\end{eqnarray}
where $\gamma$ is a positive constant making the right tail of the $P( \mathcal{N}_{sol} )$ distribution  ({\em i.e.} for values of $ \mathcal{N}_{sol} >  \langle \mathcal{N}_{sol} \rangle $)  exponentially small in $N$, as we wanted to show.
Indeed, this simple result is enough to imply that no contribution to the entropy is given by instances whose number of solutions is larger than the annealed value. The support of the probability distribution of entropy values $P(S)$ must be therefore $[0,S_{annealed}]$. In order to prove that also smaller values do not take part of the support, one would need to calculate fractional order moments, as explained in the end of next section.

\subsection{Higher order moments}
\label{app:high_mom}
For the general $n^{th}$ order momentum one can write, along the same
line of reasoning:
\begin{eqnarray}
\langle \mathcal{N}^n_{sol} \rangle &=& \sum_{\{N^{\sigma_1...\sigma_{n-1}}_o\}=0}^{N_o}
\sum_{\{N^{\sigma_1...\sigma_{n-1}}_i\}=0}^{N_i}
\frac{N_o!N_i!}{\prod_{\vec{\sigma} } N^{\sigma_1...\sigma_{n-1}}_o!N^{\sigma_1...\sigma_{n-1}}_i!} \cdot
\nonumber \\ 
& & \delta( N_o;\alpha N)\delta( N_i;(1-\alpha) N) \prod_{ \vec{\sigma} }T(\vec{\sigma}) \cdot \nonumber \\
& & \delta( N_o; \sum_{\vec{\sigma}}N^{\sigma_1...\sigma_{n-1}}_o)
\delta( N_i;\sum_{\vec{\sigma}}N^{\sigma_1...\sigma_{n-1}}_i) \\
\label{eeq7}
\end{eqnarray}
with
\begin{eqnarray}
T(\vec{\sigma}) &=& \left[
\sum_{\vec{\sigma^{(1)}}}\sum_{\vec{\sigma^{(2)}} } {\cal
P}(\sigma_1^{(1)}\sigma_1^{(2)},...,\sigma_{n-1}^{(1)}\sigma_{n-1}^{(2)}|\sigma_1,...,\sigma_{n-1})
\cdot \right. \\ &
&\left. g(\sigma_1^{(1)}\sigma_1^{(2)},...,\sigma_{n-1}^{(1)}\sigma_{n-1}^{(2)}|\sigma_1,...,\sigma_{n-1})
\right]^{N^{\sigma_1...\sigma_{n-1}}_o} \nonumber
\label{eeq7b}
\end{eqnarray}
where we are summing over all configurations overlaps with
$N^{\sigma_1...\sigma_{n-1}}_o/N^{\sigma_1...\sigma_{n-1}}_i$
output/input variables of signs $\sigma_1...\sigma_{n-1}$, the ${\cal
P}$ represent the probability of finding $n-1$ real replicas of an
input variables couple in a given function, and $g$ the value of
the $(n-1)^{th}$ averaged product of the replicated Boolean function.

As before, from the leading terms of eq.(\ref{eeq7}), one gets
\begin{eqnarray}
\langle \mathcal{N}^n_{sol} \rangle &\to& I^{(n)}(\alpha,X) \nonumber \\
I^{(n)}(\alpha,X) &=& 2^{(1-\alpha)N}
\int_0^1\prod_{{\vec \sigma}}d\{b_o\}d\{b_i\}
K^{(n)}(\alpha,X,\{ b_o,b_i \})
e^{N {\cal F}^{(n)}[\alpha,X,\{ b_o,b_i \}]} \nonumber
\end{eqnarray}
with
\begin{eqnarray}
{\cal F}^{(n)}[\alpha,X,\{ b_o,b_i \}] &=& (1-\alpha) H^{(n)} + 
\alpha D^{(n)}_{KL} (\{b_o \} | G^{(n)}) \nonumber \\
H^{(n)}(\alpha,X,\{ b_o,b_i \}) &=& \sum_{{\vec \sigma}} 
b_i^{\sigma_1...\sigma_{n-1}} \log(b_i^{\sigma_1...\sigma_{n-1}}) \nonumber \\
D^{(n)}_{KL}( \{b_o \} | G^{(n)} ) &=&
\sum_{{\vec \sigma}} b_o^{\sigma_1...\sigma{n-1}} \log\left(\frac{G^{(n)}}{b_0^{\sigma_1...\sigma{n-1}}}
\right) 
\label{eeq9}
\end{eqnarray}
with $b_{o,i}^{\sigma_1...\sigma_{n-1}} =
N_{o,i}^{\sigma_1...\sigma_{n-1}}/N_{o,i}$ and $\{ b_{o,i} \} = \{
b_{o,i}^{\sigma_1,...,\sigma_{n-1}} \}_{\vec{\sigma} \in \{ \pm 1
\}^{n-1}}$ . The explicit form of functions $G^{(n)}$ and $K^{(n)}$ is
not given here for brevity.  Due to the general form of the exponent,
the unique saddle point can always be computed as the one fulfilling
conditions:
\begin{eqnarray}
{\tilde b}_i^{\sigma_1...\sigma_{n-1}} &=& \frac{1}{2^{n-1}} \nonumber \\
{\tilde b}_o^{\sigma_1...\sigma_{n-1}} &=& G^{(n)}(\alpha,X,\{{\tilde b}_o^{\sigma_1...\sigma_{n-1}}\},
\{{\tilde b}_i^{\sigma_1...\sigma_{n-1}}\})
\label{eqsaddle2}
\end{eqnarray}
such that 
\begin{equation}
I^{(n)}(\alpha,X) = 2^{ n(1-\alpha)N} C^{(n)}(\alpha,X) = \langle \mathcal{N}_{sol} \rangle^n C^{(n)}(\alpha,X)
\label{eq:in}
\end{equation}
Unfortunately, the explicit calculation of the function
$C^{(n)}(\alpha,X)$ for all $n$ seems not to be easily accessible. Let us just finally note that one could prove that the distribution of the logarithm of the number of solutions
$P(S)$ tends to $\delta(S-S_{\mathrm{annealed}})$ in the large $N$ one by showing that:
\begin{equation}
\lim_{n\rightarrow 0} \lim_{N \rightarrow \infty} C^{(n)}( \alpha , X ) = 1
\end{equation}
which seems reasonable given the structure of eqs.~(\ref{eeq9}, \ref{eqsaddle2}, \ref{eq:in}), and the numerical result displayed in section \ref{sec:entropy}, but, again,  too difficult to be shown explicitly.
\\

\end{document}